\documentclass[aps,pof,floatfix,twocolumn,superscriptaddress]{revtex4} 
\usepackage[latin1]{inputenc}
\usepackage[english]{babel}
\usepackage[T1]{fontenc}

\usepackage{textcomp}
\usepackage{graphicx}
\usepackage{color}

\newcommand{\vk}{von~K\'arm\'an~}

\def\ADD#1{{\textcolor{black}{#1}}}  % added text
\begin{document}

\title{Stochastic dynamics of particles trapped in turbulent flows.}

\author{N.~Machicoane}
\affiliation{Laboratoire de Physique, ENS de Lyon, UMR CNRS 5672, Universit\'e de Lyon, France}
\author{M.~L\'opez-Caballero}
\affiliation{Departamento de F\'isica y Matem\'atica Aplicada, University of Navarra, P.O. Box 177, E-31080 Pamplona, Spain}
\author{L.~Fiabane}
\affiliation{Laboratoire de Physique, ENS de Lyon, UMR CNRS 5672, Universit\'e de Lyon, France}
\affiliation{Irstea, UR TERE, F-35044 Rennes, France}
\author{J-F.~Pinton}
\affiliation{Laboratoire de Physique, ENS de Lyon, UMR CNRS 5672, Universit\'e de Lyon, France}
\author{M.~Bourgoin}
\affiliation{Laboratoire de Physique, ENS de Lyon, UMR CNRS 5672, Universit\'e de Lyon, France}
\affiliation{Laboratoire des \'Ecoulements G\'eophysiques et Industriels, CNRS/UJF/G-INP UMR 5519, BP53, F-38041 Grenoble, France}
\author{J.~Burguete}
\affiliation{Departamento de F\'isica y Matem\'atica Aplicada, University of Navarra, P.O. Box 177, E-31080 Pamplona, Spain}
\author{R.~Volk}
\email{romain.volk@ens-lyon.fr}
\affiliation{Laboratoire de Physique, ENS de Lyon, UMR CNRS 5672, Universit\'e de Lyon, France}

\begin{abstract}
The long time dynamics of large particles trapped in two inhomogeneous turbulent shear flows is studied experimentally. 
Both flows present a common feature, a shear region that separates two colliding circulations, but with different spatial symmetries and temporal behaviors. 
Because large particles are less and less sensitive to flow fluctuations as their size increases, we observe the emergence of a slow dynamics corresponding to back-and-forth motions between two attractors, and a super-slow regime synchronized with flow reversals when they exist. 
\ADD{Such dynamics is substantially reproduced by a one dimensional stochastic model of an over-damped particle trapped in a two-well potential, forced by a colored noise.  
An extended model is also proposed that reproduces observed dynamics and trapping without potential barrier: the key ingredient is the ratio between the time scales of the noise correlation and the particle dynamics. A total agreement with experiments requires the introduction of spatially inhomogeneous fluctuations and a suited confinement strength.}
\end{abstract}


%\date{\today}
\maketitle

\section{Introduction}
Trapping, defined as the process where an entity remains locked in a given region of a dynamical space, is very common in physics, chemistry, and biology \cite{kramers1940,vankampen1981}. This problem is usually described as an activated process, where a particle trapped in a potential can cross a barrier under the action of random fluctuations. The case of white noise fluctuations has been largely studied, yielding a large set of tools that allows the analysis of these problems ({\it e.g.} Fokker-Planck equations).
However, many questions stay open concerning the influence of a colored noise, which we would like to address in this article using an experimental approach. Turbulent flows laden with particles give the opportunity to study a trapping problem with correlated fluctuations. Indeed, most real turbulent flows have a mean structure, with shear layers separating different circulations in which light inertial particle can be trapped for long times \cite{maxey1987,djeridi1999bubble,Climent2007}. A similar situation arises for large neutrally buoyant particles (namely particles with diameter typically of the order of the integral scale of the flow), which were for instance shown to be preferentially found in two attractors corresponding to low pressure and low fluctuations regions in von K\'arm\'an flows \cite{machicoane:njp2014}. This trapping phenomenon can be related to the lesser response of such large particles to the fluctuations of the flow, an effect which reduces the escape probability of the particle from these attracting regions. Strong and rare fluctuation events can nevertheless trigger the transition between trapping regions. Under the combined action of trapping and turbulent fluctuations, such particles are expected to exhibit a complex dynamics with sudden intermittent transitions between attractors, suggesting possible analogies with multi-stable stochastic systems.\\
We report experimental measurements of the dynamics of large particles tracked for very long times in two shear flows (counter-rotating \vk flows), and we show how this dynamics is affected by the different symmetries of the large scale circulations. In all situations we find the particle position is a strongly fluctuating quantity with a spectrum presenting a plateau at very low frequencies and a power law of exponent close to $-4$ at high frequency , reminiscent of the $f^{-2}$ Lagrangian velocity spectrum at inertial time scales, as reported for tracer particles \cite{mordant:njp2004,Ouellette:njp2006}. When trapping is present, we observe the emergence of a bistable dynamics characterized by a new power law of exponent $-1.7$ at intermediate frequency, corresponding to back-and-forth motions between the attractors. All statistical quantities are strongly influenced by the symmetries of the large scale flow and present a signature of its temporal dynamics. 
\ADD{Most results are recovered when drawing an analogy with the 1d dynamics of an over-damped particle trapped in a two-well potential and subjected to an exponentially correlated noise.  
An enhanced description of the experimental data is obtained using a different approach. An improved model without potential barrier is introduced, where the particle evolves in a large region, confined by lateral potential walls, where the trapping close to these walls comes from the effect of the correlated noise together with the spatial non homogeneity of the fluctuations. This model can reproduce the observed dynamics and the trapping, but a total agreement with experiments needs a suited confinement strength}.\\

\begin{figure}[h]
  \begin{center}
    \includegraphics[height=3.5cm]{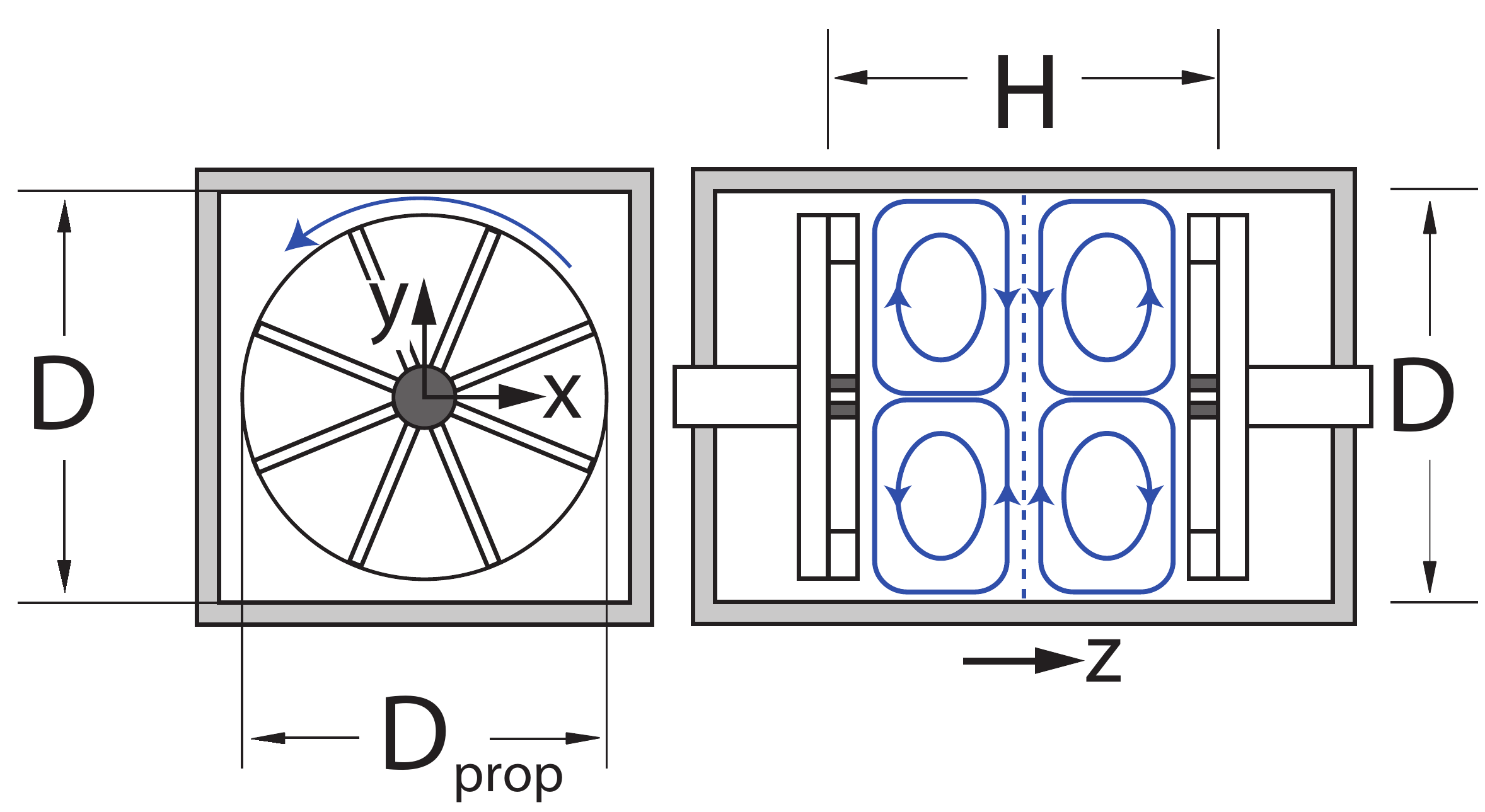}
    \includegraphics[height=3.5cm]{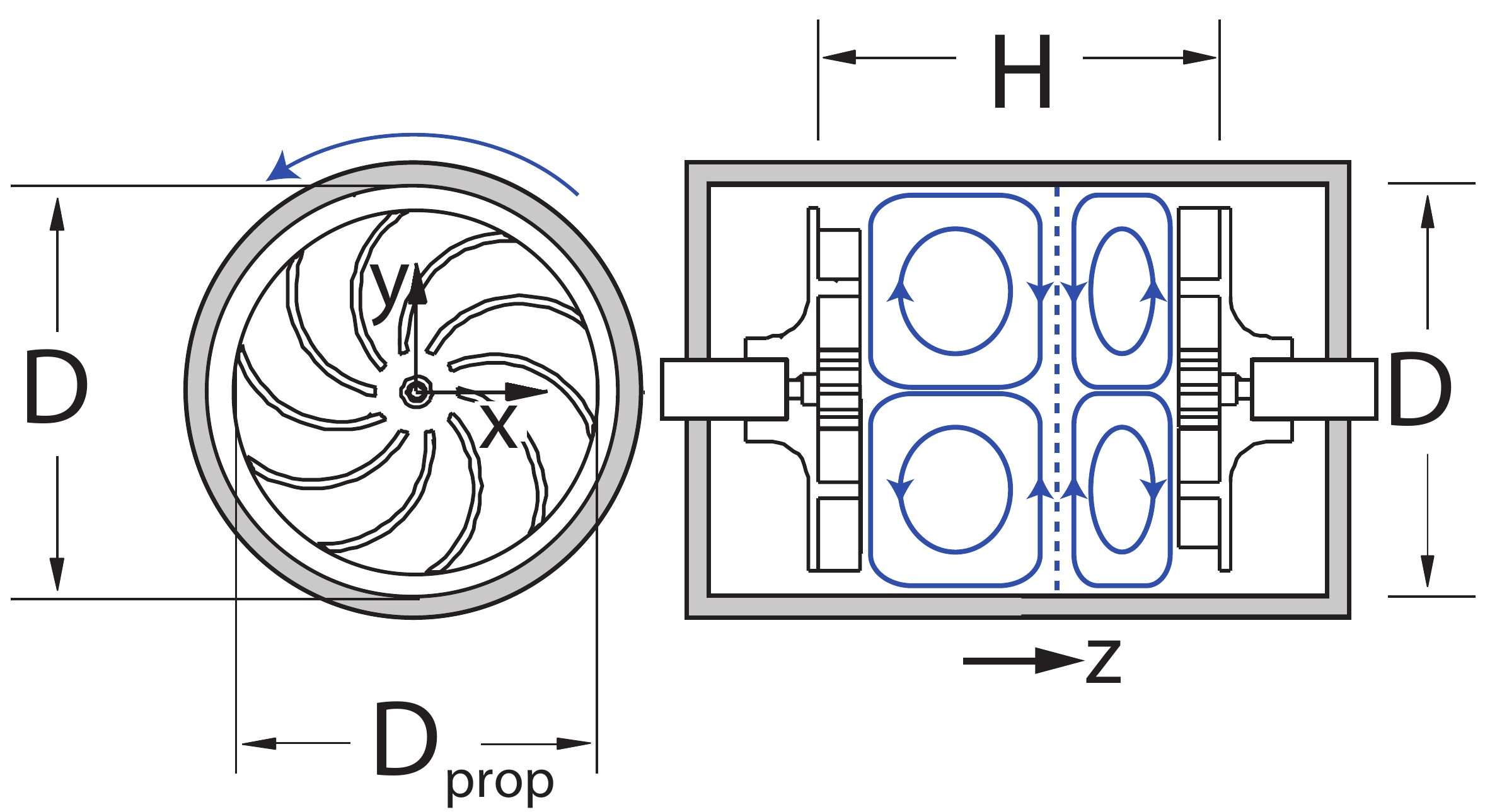}
    \caption{Sketches of the von K\'arm\'an setups and mean flows. Top: S-VK (Square-Von K\'arm\'an) configuration. Bottom: C-VK (Cylindrical-Von K\'arm\'an) configuration.}
  \label{setup}
  \end{center}
\end{figure}

%{\em Experimental Setups.---} 
\section{Experimental observations}
We investigate the long time dynamics of large particles in two water-filled von K\'arm\'an flows using two distinct setups in Pamplona and Lyon detailed in \cite{delaTorre:prl2007} and \cite{Zimmermann:prl2011}, hereafter called C-VK (Cylindrical-Von K\'arm\'an) and S-VK (Square-Von K\'arm\'an) respectively. As shown in figure \ref{setup}, both systems share common features: two coaxial propellers counter-rotating at the same frequency $f_{prop}$ are placed in a closed cavity, producing non homogeneous and non isotropic turbulent flows. Such configurations produce average flows composed of two toroidal cells, each driven by one of the propellers, with a strong shear layer in between. The two setups have disks radii $R^C_{prop}=8.75$ cm and $R^S_{prop}=9.5$ cm for the C-VK and S-VK respectively, similar distances between the disks $H^C=2R^C_{prop}=20$ cm ($H^S=2R^S_{prop}=19$ cm), and similar aspect ratios. The main difference concerns the symmetries of the cavity: the Lyon experiment uses impellers with straight blades and has a square cross section so that the mean flow of the S-VK is the most probable flow, with a shear layer in the mid plane ($z=0$) of the apparatus. On the other hand, the Pamplona experiment uses curved blades pushing the fluid with the convex side, and has a circular cross section so that the shear layer of the C-VK is always found to be displaced to one side of the apparatus. This displacement leads to an asymmetric flow with respect to the mid plane, one of the cells being bigger than the other. In addition, the shear layer can spontaneously jump to the other side of the apparatus, leading to the opposite state. These reversals occur almost instantaneously while the duration of each state (here after referred to as C$^+$ and C$^-$, when the shear layer z-position is positive and negative respectively) lasts for long times \cite{delaTorre:prl2007}, the Eulerian velocity time spectra presenting a $f^{-2}$ regime at extremely low frequencies \cite{lopez2013}. For the rotating frequencies considered in the experimental runs ($f_{prop} \in [1,4]$ Hz), both configurations produce fully turbulent flows with similarly high Reynolds numbers $Re={2 \pi R_{prop}^2 f_{prop}}/{\nu}=\mathcal{O}(10^5)$.\\%, with an integral length scale $L_{int}=16$ mm for the C-VK (resp. $30$ mm for the S-VK).
In the first set of experiments, we track polyamide spheres with density 1.14 kg.$m^{-3}$ and diameters $[6, 10, 18, 24]$ mm in the whole flow volume of the S-VK using the 3d Particle Tracking Velocimetry setup described in \cite{Zimmermann:prl2011}. In the second set of experiments, almost neutrally buoyant water-filled celluloid caps particles, with diameters $[10, 20, 30, 40]$ mm, are tracked in the entire flow of the C-VK using only one camera. Because the camera is located at a distance much larger than the dimensions of the vessel, this configuration gives the 2d position $(y,z)$ with only a weak bias. Datasets are obtained with a moderate sampling frequency ($45$ Hz for S-VK, $15$ Hz for C-VK) so that the fastest scales of particles motions are not resolved. This allows for tracking particles during very long times, of the order of thousands integral time scale $T=1/f_{prop}$, hence giving access to the particles long time dynamics.\\% Due to the lack of time resolution, trajectories are not differentiable, and we only focus on particle position PDF and spectra.\\

\begin{figure}[h]
  \begin{center}
  \includegraphics[width=\columnwidth]{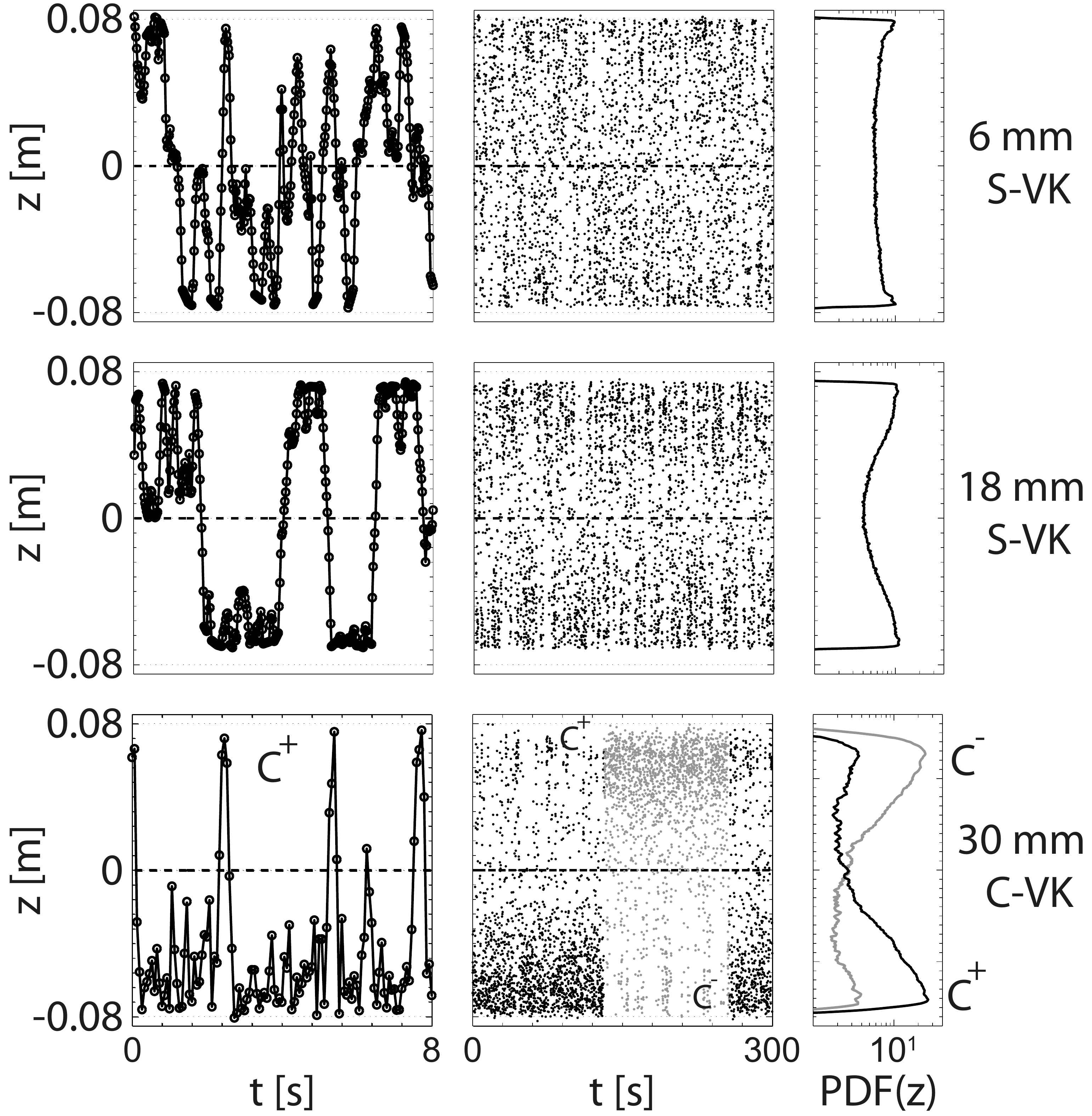}
    \caption{Time series of the axial position of a particle on short time (first column) and long time (second column) with the probability density functions (third column). The first and second lines stand for 6 and 18 mm particles freely advected in the S-VK configuration with a rotating frequency $f_{prop}=3$ Hz, while the last line is a 30 mm particle moving in the C-VK with $f_{prop}=3.16$ Hz. For the C-VK, the signal has been labelled according to the position of the shear layer corresponding to C$^+$-VK (black) and C$^-$-VK (gray) states. The PDF are conditioned on the two states C$^\pm$-VK.}
  \label{pos_sig}
  \end{center}
\end{figure}

%{\em Experimental Results.---} 
Figure \ref{pos_sig} shows time recordings of the axial position $z(t)$.
In the symmetric (S-VK) configuration, particles dynamics exhibit back-and-forth motions between two regions of equally high probability situated on the sides of the vessel. This sampling behavior, which increases with the particle diameter $D$ as already observed in \cite{machicoane:njp2014}, is well captured by the position probability density function (PDF), being almost flat for particles  smaller than $10$ mm while presenting 2 preferential regions near the disks (lumps) and a hollow region in the center for larger particles. The particles are also trapped in the asymmetric (C-VK) configuration, but with a more intermittent dynamics. On short times, the particle is most likely found on one side of the vessel (here corresponding to negative values), and is only allowed to visit the other side during very short excursions. This behavior, observed here because the shear layer was displaced toward positive values of $z$ (C$^+$ state of the flow) while this signal was taken, is confirmed when labeling the long-time signal according to the position of the shear layer. Indeed, by looking at simultaneous recordings of particle position and flow velocity measured by Laser Doppler Velocimetry, we observed the reversals in the particle position are synchronized with the reversals of the large scale flow, the particle being trapped in the largest cell \cite{mlctes}. In this asymmetric configuration, particle position PDF conditioned on the two mirror states C$^\pm$ are then asymmetric as a consequence of the global asymmetry of the large scale flow, and exchange in the symmetry $z\rightarrow -z$. For both C-VK and S-VK configurations, we can determine the time a particle is trapped in one side of the apparatus before it escapes to the other side, called escape time and noted $\Delta t$ here. For the C-VK configuration, it is necessary to take into account the position of the shear layer to dissociate an escape from, and toward, the most probable region. Whatever the setups and configurations, we always find that the escape times follow decaying exponential distributions: $PDF(\Delta t) \sim 1/T_{0}e^{-\Delta t/T_{0}}$, with $T_{0}=\left\langle \Delta t \right\rangle$. This kind of PDF distributions for escape  times, corresponding to independent random jumps, is often observed in fluctuation-driven bistable systems\cite{kramers1940,constable2000,benzi2005,berhanu2007}.\\

\begin{figure}[h]
  \begin{center}
  \includegraphics[width=\columnwidth]{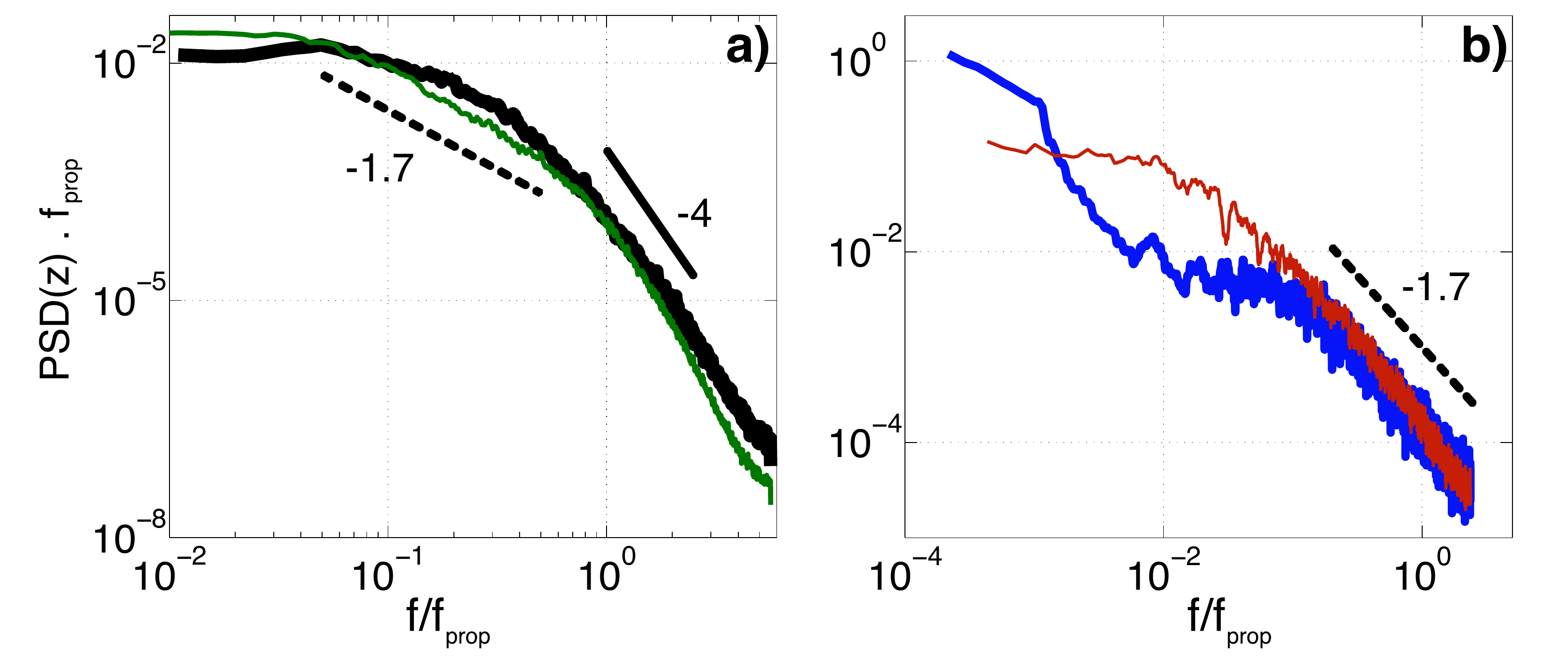}
    \caption{a) Power spectrum densities (noted PSD) of the axial position $z$ for 6 mm (thin solid line) and 18 mm (thick solid line) particles in the S-VK with $f_{prop}=3$ Hz. b) PSD of $z$ for 10 mm (thin solid line) and 30 mm (thick solid line) particles in the C-VK with $f_{prop}=3.16$ Hz.}
  \label{spectre}
  \end{center}
\end{figure}

\indent In order to better characterize what range of time scales in the particle dynamics is affected by the heterogeneous sampling, we compute the axial position power spectra of the time series displayed in figure \ref{pos_sig}, and present the results in figure \ref{spectre}. For the S-VK configuration, all spectra are monotonically decreasing and present a plateau in the very low frequency range. Such behavior corresponds to a total decorrelation of the axial position whose variance remains finite, which is expected for particles transported in a bounded domain. Concerning the $6$ mm particle, one observes a slow transition toward a power law of exponent close to (but steeper than) $-4$ in the high frequency range $f/f_{prop} \geq 1$, as would be observed for tracers in similar flows \cite{mordant:njp2004,Ouellette:njp2006}. For the very large particles considered here, which start to be trapped in low pressure, low fluctuations regions when their diameter exceeds $D \simeq 10$ mm \cite{machicoane:njp2014}, the influence of non homogeneous sampling manifests itself in the intermediate frequency range $f/f_{prop} \in [0.1, 1]$ as a new power law of exponent $\alpha \simeq -1.7$ corresponding to the long time oscillations observed in figure \ref{pos_sig}. A similar power law is observed in the C-VK for all the large particles considered here, but with a plateau whose height decreases at increasing particle diameter, indicating the position variance decreases within one (C$^+$ or C$^-$) state. The main difference with the S-VK configuration is the appearance of a super slow regime for the largest particles, for which the position of the most probable region synchronizes with the flow reversals \cite{mlctes}. 

\begin{figure}[h]
  \begin{center}
  \includegraphics[width=\columnwidth]{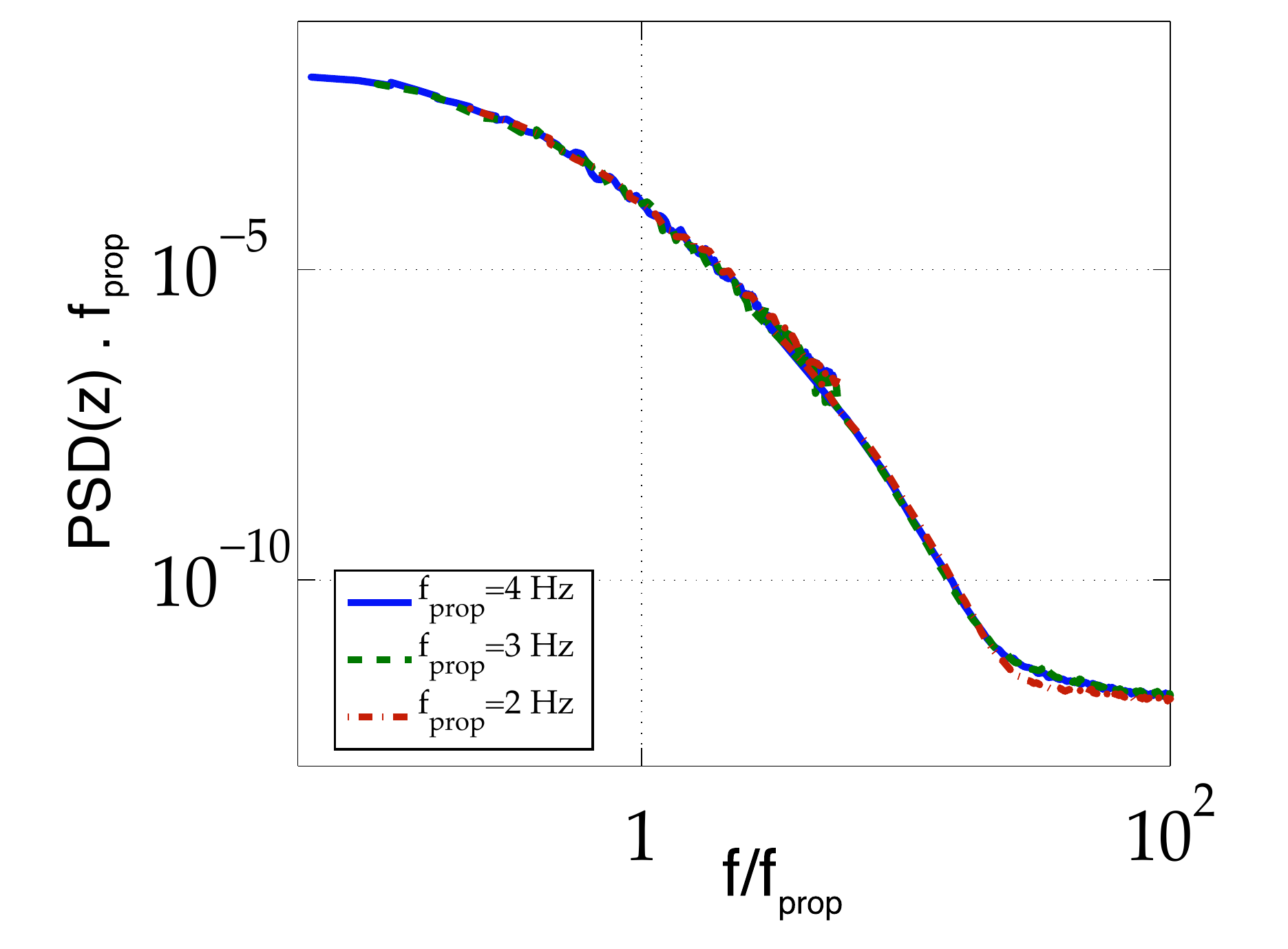}
    \caption{\ADD{Normalized spectra $f_{prop} PSD(z)$ of the axial position $z$ for 6 mm particles, plotted as a function of the reduced frequency $f/f_{prop}$, measured for $3$ rotating frequencies $f_{prop}=2,3,4$ Hz in the S-VK setup. $f_{prop}=4$Hz (solid line),  $f_{prop}=3$Hz (dashed line), $f_{prop}=2$Hz (dashed-dotted line).}}
  \label{spectre6mm}
  \end{center}
\end{figure}

%\DEL{All these observations are robust when varying the Reynolds number by changing the frequency of the propellers. For any particle size in the range $D\in[6-24]$mm, we observed that the normalized frequency spectra $f_{prop} PSD(z)$ reduce to a single curve when plotted as a function of reduced frequency $f/f_{prop}$. This result is valid providing the turbulence is fully developed and the particle diameters are of the order of the flow integral scale.} 

\ADD{All these observations are robust when varying the Reynolds number by changing the frequency of the propellers. As demonstrated in figure \ref{spectre6mm}, which displays results obtained in the case of the smallest ($6$ mm) particles advected in the S-VK setup, the normalized frequency spectra $f_{prop} PSD(z)$ reduce to a single curve when plotted as a function of reduced frequency $f/f_{prop}$. Note that this figure presents a combination of $6$ curves, $3$ corresponding to very long runs with a sampling frequency $45$ Hz while the $3$ others correspond to short and well resolved trajectories recorded at $3000$ Hz. Such collapse of the spectra from very low up to very high frequencies was observed to hold for any particle size in the range $D\in[6-24]$ mm. It is valid providing the turbulence is fully developed and the particle diameters are of the order of the flow integral scale.} 

\section{Stochastic models}

For both setups the dynamics in the $z$ direction is determined by the trapping of the spheres around two very well defined $z$-positions, close to the propellers. We propose to describe this phenomenon as an over damped particle moving in a potential defined inside the experimental cell, and driven by random fluctuations: 

\begin{equation} 
\frac{dz}{dt}=-\frac{dV}{dz}+v(z,t), 
\end{equation} 

where $V$ is a (non dimensional) potential accounting for confinement and eventually trapping, and $v$ is the Lagrangian fluctuating velocity the particle would have in an unbounded domain. Because turbulence intensity varies in space, we write it $v(z,t) = B_0\,s(z)\,\eta(t)$, where $B_0$ is the fluctuations intensity, $s(z)$ a spatial profile, and $\eta(t)$ a random variable. Very small scales of turbulence are not expected to drive long time dynamics, we thus only retain inertial range fluctuations which are modeled as an Ornstein-Uhlenbeck process with correlation time $\tau_p$ \cite{snyder1971,sato1987,Yeung1989,mordant:njp2004}. 
\ADD{
\begin{equation} 
\frac{d\eta}{dt}=-\frac{1}{\tau_p} \eta + \sqrt{\frac{2 \eta_0^2}{\tau_p}} \xi(t),
\end{equation} 
where $\eta_0=1$, and $\xi(t)$ is a Gaussian noise satisfying $\langle\xi(t)\xi(t')\rangle=\delta(t-t')$.} Particles may avoid the central region of the flow because of the mean flow topology, or because turbulence is stronger in this region. To account for these two possibilities, we used two different models. 

\subsection{Double-well model}

In the first model, the {\em Double-well model},  the fluctuations are homogeneous ($s(z)=1$), and the potential presents two wells with lateral walls of constant stiffness: 

\begin{eqnarray}
& V(z)=\delta(\frac{\displaystyle z^{4}}{\displaystyle 4}-\frac{\displaystyle z^{2}}{\displaystyle 2})-\lambda z \hspace{1cm} |z|<1 \\
& V(z)=4(\frac{\displaystyle z^{4}}{\displaystyle 4}-\frac{\displaystyle z^{2}}{\displaystyle 2}) - \lambda z \hspace{1cm} |z|\geq1.
 \end{eqnarray}
 
\begin{figure}[h]
  \begin{center}
      \includegraphics[width=\columnwidth]{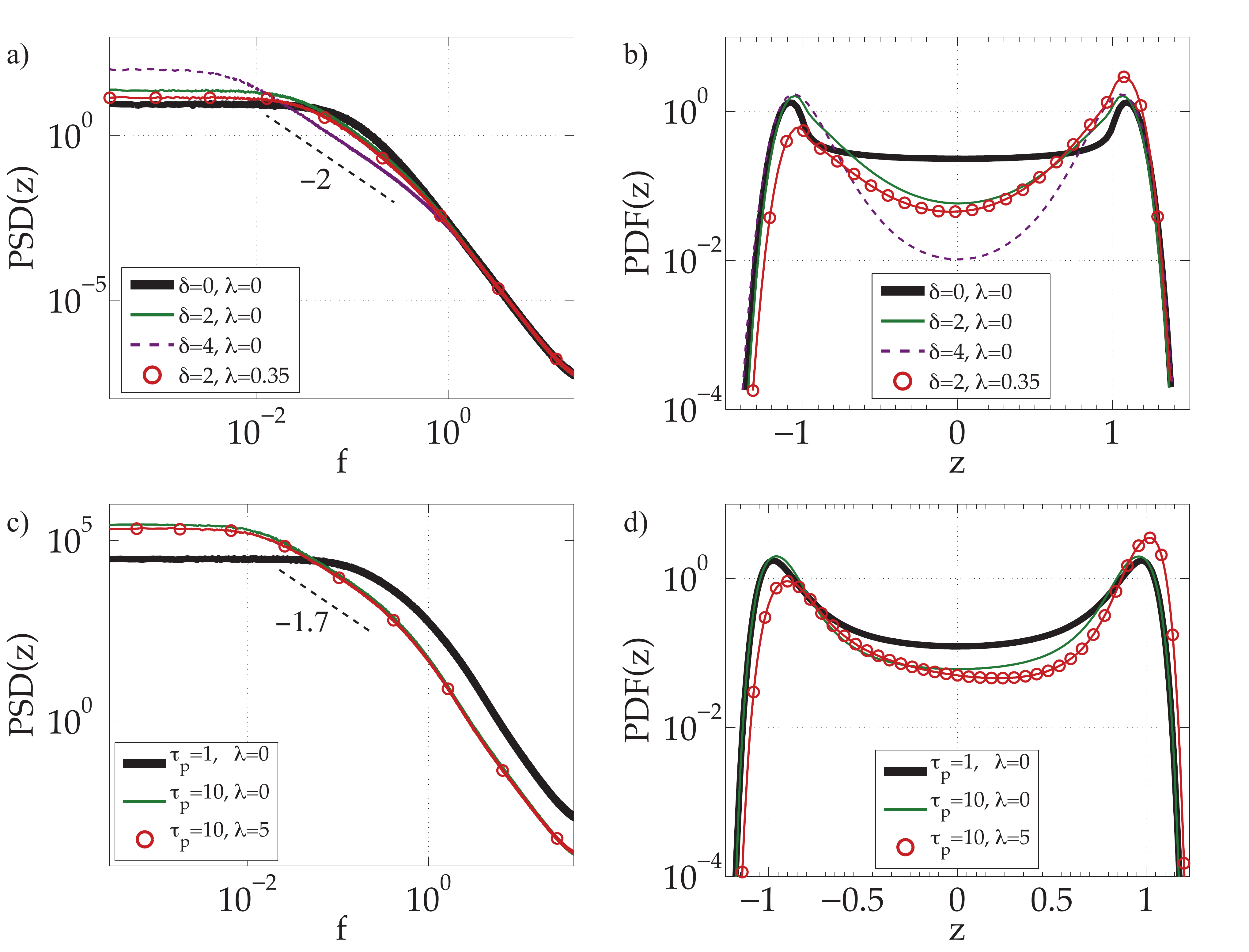}
    \caption{\ADD{a) Position PSD for the {\em Double-well model}. Thick solid: reference case $(\delta,\lambda)=(0,0)$. Thin solid: symmetric case $(\delta,\lambda)=(2,0)$. Dashed line: symmetric case, $(\delta,\lambda)=(4,0)$. Open circles: asymmetric case $(\delta,\lambda)=(2,0.35)$. The dotted line indicates a power of exponent $-2$. b) Corresponding position PDF for the {\em Double-well model} with conventions of figure a). c,d) Position PSD and PDF for the {\em Non homogeneous model}. Thick solid: reference case $(\tau_p,\lambda)=(1,0)$. Thin solid: symmetric case $(\tau_p,\lambda)=(10,0)$. Open circles: asymmetric case $(\tau_p,\lambda)=(10,5)$}.}   \label{model}
  \end{center}
\end{figure}

The shape of the potential is then set by the couple of parameters $(\delta,\lambda)$, which control the height of the barrier between the wells  and the asymmetry observed in each state C$^\pm$ of the C-VK configuration. This kind of stochastic model is known to reproduce the behavior of bistable systems such as magnetic field reversals \cite{berhanu2007,constable2000} or turbulent flow reversals \cite{benzi2005,delaTorre:prl2007}. In this model, the correlation time and amplitude of fluctuations are chosen so that $\sqrt{B_0} \tau_p$ equals a significant fraction of the box size to prevent turbulent diffusion when $(\delta,\lambda)=(0,0)$. \ADD{Figure \ref{model} (a,b) present the position spectra and PDF obtained for 3 cases of the {\em Double well model} with the choice $\tau_p=1$ and $B_0=1$: confinement only (thick solid line, $\delta=\lambda=0$), symmetric double-well potential (thin solid line, $\delta=2$, $\lambda=0$; dashed line, $\delta=4$, $\lambda=0$) and asymmetric double-well potential (open circles, $\delta=2$, $\lambda=0.35$).} Without any barrier, the sampling is almost  homogeneous with increased probability on the sides, as already observed for the $6$ mm particles. This is due to the finite correlation time $\tau_p$, and leads to a spectrum with a low-frequency plateau and a power law of exponent $-4$ in the high frequency range, resembling the dynamics of the smallest particles in the S-VK. When adding a barrier, we observe a strong preferential sampling with two trapping regions around $z=\pm1$, leading to a power law of exponent $-2$ for intermediate frequencies, similar to the dynamics of large particles in the S-VK. \ADD{A change of the height of the barrier $\delta$ modifies the depth of the central valley of the PDF distribution from one to two orders of magnitude compared to the maxima, as well as the extent of the -2 slope in the intermediate frequency region. However, a preferential trapping stronger than the one found in the experiments seems to be necessary to recover this slope on a comparable frequency range.}

Adding an asymmetry to reproduce the C$^-$ state, we obtain a PDF and spectrum similar to those of C-VK, with decreased height of the plateau at increasing asymmetry as observed in the experiment when $D$ increases. This indicates that if increasing particle size corresponds to an increase of the barrier height in the symmetric configuration, it corresponds to an increased asymmetry when the mean flow is asymmetric. For all set of parameters, we observed exponential distributions of escape times, corresponding to independent random jumps with transition times much shorter than residence time.
\ADD{This scale separation between escape and transition times is much less evident in the experiment, and can be one of the reasons why this model cannot capture the whole range of experimental dynamical features. This {\em Double well model} relies on the activation of the particle. When activated, the particle transits rapidly from one
potential well to the other.  A model relying on a different approach could reproduce the position spectra exponent $-1.7$ and yields an even better agreement with the trapping dynamics.}

\subsection{Non homogeneous model}
\label{nhmgmodel}
\ADD{With this idea in mind, we replaced the double well potential with a wide and flat potential that accounts only for the confinement between the propellers. Providing the potential is not quadratic, trapping is obtained by adjusting the ratio between the time scales for the particle dynamics $\tau_p/T$, where $T$ is defined with the deterministic equation $\dot{z}=-dV/dz$ and a typical length-scale $\sqrt{\langle z^2 \rangle}$, and $\tau_p$ is the correlation time of the fluctuations~\footnote{\label{note1}\ADD{Using a quadratic potential (case $n=2$) would lead to a two-dimensional Ornstein-Uhlenbeck process for the variables $(z,\eta)$ \cite{vankampen1981}. In such a case the joint probability $P(z,\eta,t)$ converges toward a Gaussian stationary solution $P_s(z,\eta) \propto \exp(-az^2+bz\eta-c\eta^2)$ whatever the correlation time $\tau_p$ if $B(z)$ is constant, or if the non homogeneity is too weak as compared to the confinement. We therefore only consider potentials with exponents $n \geq 4$ leading to nonlinear Langevin equations. For such non linear equations, increasing the correlation time $\tau_p$ leads to bimodal position PDFs.}}. When these times are similar there is no trapping, but when the fluctuations correlation time is large, it induces the trapping as a mixed effect of a noise driven force that pushes the particle towards one of the propellers on larger time-scale compared to the time scales of the particle dynamics. Other features that appear in the experiments are also included, as the non homogeneity of turbulence intensity or the stiffness of the lateral potential walls. Although these parameters allow a fine tuning of the final results, they are not of crucial importance to induce the trapping. We would like to highlight that this model induces the trapping even {\em without} energy barriers.} %%

For this {\em Non homogeneous model}, the confinement potential is: 
\begin{equation} V(z)=z^n/n-\lambda z, \end{equation} where $n \geq 4$ is an even integer and $\lambda$ the asymmetry parameter (see footnote in the references). Because fluctuations are stronger close to the plane $z=0$ in the experiments, we set $s(z)= \exp ( -z^2 / 2 \sigma_s^2)$, where $\sigma_s$ is the typical scale of turbulence non homogeneity (equal to a fraction of the system size in the experiment). For the results presented here, we set $B_0=10$, $\sigma_s=1$ and $n=8$, but the precise values of the first two parameters were found to have a weak impact on the results. The only remaining parameters to be varied are then the asymmetry parameter $\lambda$ and the particles velocity correlation time $\tau_p$,  a quantity which increases with the particle diameter because large particles are only sensitive to eddies larger than their own size \cite{Qureshi:prl2007,Volk:jfm2011,Calzavarini2012}. We would like to stress here that no potential barrier at $z=0$ is imposed in this second approach, nevertheless the system is able to reproduce trapping close to the propellers when increasing the particles correlation time $\tau_p$ as demonstrated in figures \ref{model} (c,d). A particle with a small correlation time ($\tau_p=1, \lambda=0$) has a dynamics similar to the one observed for small particles in the S-VK, while a larger particle with a larger correlation time ($\tau_p=10, \lambda=0$) has a bistable dynamics with a position power spectrum presenting again three regimes, but with an exponent $\beta \sim -1.7$ at intermediate frequencies, closer to the experimental result. \ADD{Concerning the influence of an asymmetry, the {\em Non homogeneous model} with $\tau_p=10, \lambda=5$} gives similar results  as those obtained with the {\em Double well model}: the height of the plateau at low frequency decreases, corresponding to a reduced rms value for the particle position. However we observed that the {\em Non homogeneous model} yields different exponents in the range $[-2.2,-1.5]$ when varying the stiffness of the lateral walls\ADD{, a consequence of the non linearity of the restoring force}. %, which is not the case with the {\em Double well model}.
We may then deduce that the precise value of the exponent in the experimental results has (at least partially) its origin in the confinement strength in the flow combined with the non homogeneity of the turbulence intensity.\\%of the potential.\\

%{\em Conclusion.---} 
\section{Conclusion}
We have studied the dynamics of large particles in two turbulent flows presenting a large scale structure with recirculating regions. Because large particles are less sensitive to flow fluctuations as their diameter increases, they tend to be trapped in attractors close to the disks corresponding to low pressure, low fluctuations regions of the flow, and exhibit a complex bistable dynamics under the combined action of the mean flow topology and non homogeneous turbulence. This dynamics is very sensitive to any eventual asymmetry of the mean flow structure, and presents the main features of a stochastic bistable system. In order to study the influence of the different mechanisms at play, we described the dynamics using two different models. We first considered a classical {\em Double well model} consisting of an over damped particle moving in a two-well potential under the action of fluctuations with exponential correlation. 
%\DEL{When varying the height of the barrier, this model reproduces most of the experimental results, and demonstrates the influence of the noise correlation, at the origin of an increase of the position PDF close to the disks even for the smallest particles.}
\ADD{This model reproduces most of the experimental results when varying the height of the barrier. It also demonstrates the influence of the noise correlation, at the origin of an increase of the position PDF close to the disks even for the smallest particles. However, the model only predicts position spectra with a power law of exponent $-2$ at intermediate frequency corresponding to sharp uncorrelated reversals of particle position.\\
Using a second model based on the different time-scales between the particle dynamics and the noise correlation we can reproduce the trapping without any potential barrier that separates two spatial regions. While it was found that the ratio between the time-scales of the noise and of the particles dynamics is the most important parameter for the trapping, we observed the back and forth dynamics is influenced by non-homogeneities and the confinement strength. Indeed, the value of the slope of the position power spectra in the intermediate frequency regime is shown to be finely tuned by both parameters.}
Finally, we note that most real turbulent flows present a mean structure with recirculating regions and non homogenous fluctuation fields where inertial particles, such as bubbles, may be trapped for long times \cite{djeridi1999bubble,Climent2007}. 
We expect that these particles also present a complex long time dynamics corresponding to jumps between attractors. Using an approach similar to the one presented in this article, it would be possible to derive a model describing their dynamics, whose key characteristics are the trapping potential and the temporal and spatial distributions of the driving noise.\\

{\bf Acknowledgments} %The authors want to thank Freddy Bouchet for stimulating discussions. 
N. Machicoane and M.~L\'opez-Caballero contributed equally to this work. This work is supported by French research programs ANR-12-BS09-0011, ANR-13-BS09-0009 and Projet Emergent PALSE/2013/26, and Spanish research programs FIS2011-24642 and FIS2014-54101-P.\\

%\bibliography{main}
\bibliography{biblio_Lsphere}
\end{document}